\documentclass[twocolumn, pra, showpacs,superscriptaddress]{revtex4}
\usepackage{amssymb}
\usepackage{graphicx}
\usepackage{dcolumn}
\usepackage{bm}
\usepackage{amsmath}

\begin{document}

\title{Spectroscopy and spin dynamics for strongly interacting few spinor
bosons in one-dimensional traps}
\author{Yanxia Liu}
\affiliation{Institute of Theoretical Physics, Shanxi University, Taiyuan 030006, China}
\author{Shu Chen}
\email{schen@iphy.ac.cn}
\affiliation{Beijing National Laboratory for Condensed Matter Physics, Institute of
Physics, Chinese Academy of Sciences, Beijing, 100190, China}
\affiliation{School of Physical Sciences, University of Chinese Academy of Sciences,
Beijing, 100049, China}
\affiliation{Collaborative Innovation Center of Quantum Matter, Beijing, China}
\author{Yunbo Zhang}
\email{ybzhang@sxu.edu.cn}
\affiliation{Institute of Theoretical Physics, Shanxi University, Taiyuan 030006, China}

\begin{abstract}
We consider a one-dimensional trapped gas of strongly interacting few spin-1
atoms which can be described by an effective spin chain Hamiltonian. Away
from the $SU(3)$ integrable point, where the energy spectrum is highly
degenerate, the rules of ordering and crossing of the energy levels and the
symmetry of the eigenstates in the regime of large but finite repulsion have
been elucidated. We study the spin-mixing dynamics which is shown to be very
sensitive to the ratio between the two channel interactions $g_{0}/g_{2}$
and the effective spin chain transfers the quantum states more perfectly
than the\ Heisenberg bilinear-biquadratic spin chain.
\end{abstract}

\maketitle

\section{Introduction}

Spinor quantum gases have attracted a lot of attentions due to their rich
physics in quantum coherence, spin dynamics \cite%
{Kurn2013,Law,Chang,Naylor,Zhang}, long-range order \cite{Yi,Santos},
quantum magnetism \cite{Campbell} and symmetry breaking \cite{Navon}. The
majority of researches in many-body system involved trapped spinor atoms
such as $^{23}$Na and $^{87}$Rb experimentally realized in many cold atom
labs \cite{Stamper,Liu2009,Vengalattore,Jacob,Vinit,Fukuhara}. Among these,
spin $s=1$ system plays a central role in the fundamental understanding of
topological quantum phase transition of condensed materials and in modern
technologies including, for instance, data storage \cite{Morozov}, spin
currents \cite{Sinova}, spin vortex \cite{Yi,Li}, etc. In low-dimensional
system, owing to the liberation of the spin degrees of freedom, a major
focus is the understanding of quantum magnetism of higher spin, which have
their origin in the underlying microscopic processes between elementary
spins.

Recently the experiment in Heidelberg showed strong evidences that the spin
chain of few cold atoms in one-dimensional system without an underlying
lattice can be realized in vicinity of a scattering resonance \cite%
{Murmann2015}. On the theoretical side, the general form of the effective
spin-chain model for strongly interacting atomic gases with an arbitrary
spin in the one-dimensional(1D) traps has been presented \cite%
{Deuretzbacher2008,Deuretzbacher,Yang,Yang1,Yang2,Lindgren,Volosniev}. This
provides a platform for the research of basic magnetic processes in few-body
system. For the two-component system with a large but finite strong $s$-wave
interaction, the transition between the ferromagnetism (FM) and
anti-ferromagnetism (AFM) phase, the theorem on the level crossing between
singlet ground state and the maximum spin state have been studied \cite%
{Cui,Hu,Guan,Lieb}. In addition, the effect of a weak additional $p$-wave
interaction on the magnetic orders of the ground state has already been
addressed \cite{Yang2015,Hu1}. Due to the existence of two-channel
interaction, the spin-1 system exhibits much richer phenomena than the
two-component system, which is expected to give rise to rich magnetic
properties in the strongly interacting limit.

Based on the effective spin-chain model to strongly interacting trapped
boson gases with spin-1, we study the properties of the magnetic ground
state in the strongly repulsive regime and explore the ordering and crossing
of the energy levels when the interaction transfers from the FM to AFM. By
analyzing the symmetry of system in the presence of a spin-dependent
magnetic gradient and a transverse magnetic field, we show how the ground
state may be manipulated and the atoms would collide in the spin-mixing
dynamics depending mainly on the ratio between of interaction strengths in
the two collisional channel. Spin-1 systems, in comparison to spin-1/2
systems, offer a better security for encoding and transferring quantum
information, primarily due to their larger Hilbert spaces. We further study
the quantum state transfer (QST) between the two ends of the spin chain. The
effective spin-1 chain provides a site-dependent spin-coupling protocol, due
to background trap potential and the resultant inhomogeneous particle
density. The effective spin-chain coupling protocol is expected to show more
beneficial features than the\ Heisenberg bilinear-biquadratic\ (HBB)
spin-chain coupling.

The rest of the paper is organized as follows. In Section II, we give the
detailed spectrum and eigenstates of the effective spin-1 chain model. In
Sec. III, based on the effective spin-chain model, we study the
spin-changing dynamics and the efficiency and advantage of the QST with this
effective spin chain. Finally we conclude in section IV.

\section{Spectroscopy}

\begin{figure*}[tbp]
\includegraphics[width=0.9\textwidth]{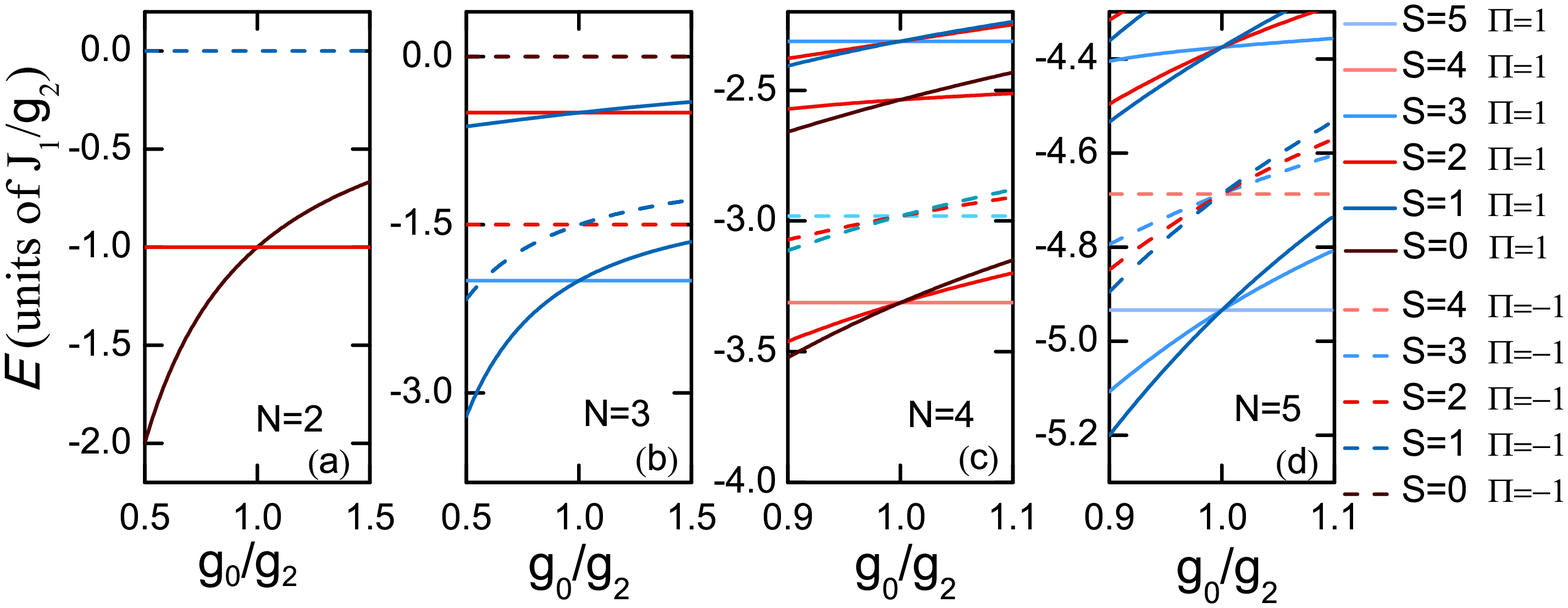}
\caption{Energy spectrum of $H_{eff}$ of $N=2,3,4,5$ particles with total
spin $S$ as a function of $g_{0}/g_{2}$. To get a closer look at the law of
the energy level crossing, we zoom into the area near $g_{0}/g_{2}=1$. On
the side of $g_{0}/g_{2}<1$, $E_{S}^{1}$ increases with total spin $S$, and
the ground state is a singlet $S=0$ for $N=2,4$ and a triplet $S=1$ for $%
N=3,5$ in the lowest bunch. On the side of $g_{0}/g_{2}>1$, $E_{S}^{1}$
increases with decreasing total spin $S$, and the ground state is FM with $%
S=N$ in the lowest bunch.}
\label{fig1}
\end{figure*}

We consider $N$ interacting atoms of hyperfine spin $s=1$ and mass $M$ in a
one-dimensional harmonic trap. The system Hamiltonian is
\begin{equation}
H=\sum_{i=1}^{N}h\left( x_{i}\right) +\sum_{i<j}^{N}\left( c_{0}+c_{2}%
\mathbf{s}_{i}\cdot \mathbf{s}_{j}\right) \delta \left( x_{i}-x_{j}\right) ,
\label{H}
\end{equation}%
where $\mathbf{s}_{i}=\left( s_{i}^{x},s_{i}^{y},s_{i}^{z}\right) $ is the
spin-1 matrix for the $i$-th atom, $c_{0}=\left( g_{0}+2g_{2}\right) /3$ and
$c_{2}=\left( g_{2}-g_{0}\right) /3$ with $g_{0}$ and $g_{2}$ the coupling
constants in the scattering channels with total spin $S=0$ and $2$,
respectively \cite{Ho,Ohmi}. For $F=1$ $^{87}$Rb, both the sign and the
magnitude of $c_{2}$ and hence the magnetic nature of the system can be
altered significantly in a relatively wide rang by means of a high 
resolution photoassociation
spectroscopy of the atoms to some excited molecular states as shown in
Chapman's experiment \cite{Hamley}. On the other hand, it has been proposed
that a multichannel scattering resonance can be achieved for spinor bosons
confined in one-dimension geometry with an additional spin-flipping rf field
and the interaction in the two channels $S=0$ and $2$ in our system can be
tuned to be large simultaneously near the resonance \cite{Cui2014}. In the
single-particle Hamiltonian
\begin{equation}
h\left( x\right) =-\frac{\hbar ^{2}}{2M}\frac{d^{2}}{dx^{2}}+\frac{1}{2}%
M\omega ^{2}x^{2}-Gxs^{z}+\Omega s^{x},  \label{h}
\end{equation}%
with $\omega $ being the trapping frequency, each particle feels a
transverse magnetic field of strength $\Omega $ and a spin-dependent
magnetic gradient of strength $G$, which are extremely small perturbations
and will not cause any noticeable effects in a weakly interacting system
\cite{Cui}. Note that both $\Omega $ and $G$ have absorbed in them the Land%
\'{e} factor $g_{s}$ and the Bohr magneton $\mu _{B}$. This will change a
lot in the strongly interacting system as can be seen in this study.

Notably, in the absence of the external field, the eigenfunctions of the
few-particle system have been exactly solved \cite%
{Girardeau1,Girardeau2,Deuretzbacher} by means of the Bose-Fermi mapping in
the Tonks-Girardeau limit. The ground state wave function of a spin-1 system
with infinite interaction is described by \cite{Deuretzbacher,Yang,Yang1}
\begin{eqnarray}
&&\Psi \left( x_{1},s_{1};\cdots x_{N},s_{N}\right) \\
&=&|\phi _{F}\left( x_{1}\cdots x_{N}\right) |\sum_{P}P\left[ \theta \left(
x_{1}\cdots x_{N}\right) \chi \left( s_{1}\cdots s_{N}\right) \right] ,
\notag
\end{eqnarray}%
where $\theta \left( x_{1}\cdots x_{N}\right) =1$ if $x_{1}\leq \cdots \leq
x_{N}$ and zero otherwise, $x_{i}$ and $s_{i}=1,0,-1$ are position and spin
indices of the $i$-th particle, respectively. The wave function $\phi _{F}$
is taken as the ground state of $N$ spinless fermions, i.e. the Slater
determinant made up of the lowest $N$-level of eigenstates, while the spin
wave function $\left\vert \chi \right\rangle $ can be written as a
superposition of spin Fock states $\left\vert m_{1}m_{2}\cdots
m_{N}\right\rangle $, which means the $i$th spin is in the $m_{i}$ state,
i.e. $\left\vert m_{i}\right\rangle =\delta _{s_{i}m_{i}}$. The permutation $%
P$ acts on both the spatial and spin wave functions and ensures the symmetry
upon particle exchange. The model in the regime of large but finite
repulsion can be mapped to an effective ferromagnetic chain of spin-1 bosons
to the first order of $g_{0}^{-1},g_{2}^{-1}$ \cite{Yang}
\begin{equation}
H_{eff}=-\sum_{i=1}^{N-1}J_{i}\left( \frac{1}{g_{0}}P_{0}\left( i,i+1\right)
+\frac{1}{g_{2}}P_{2}\left( i,i+1\right) \right) .  \label{he}
\end{equation}%
Instead of representing the model in terms of the permutation operators $%
P_{ij}$ of neighboring spins \cite{Deuretzbacher}, we here classify the
states according to the collisional channels of total spin $S$ of the two
sites. For spin-1 atoms we define the projection operators in the total spin
$S=0$ and $S=2$ channels as
\begin{equation}
P_{0}\left( i,i+1\right) =\frac{\left( \mathbf{s}_{i}\cdot \mathbf{s}%
_{i+1}\right) ^{2}-1}{3}
\end{equation}%
and%
\begin{equation}
P_{2}\left( i,i+1\right) =\frac{\left( \mathbf{s}_{i}\cdot \mathbf{s}%
_{i+1}\right) ^{2}}{6}+\frac{\mathbf{s}_{i}\cdot \mathbf{s}_{i+1}}{2}+\frac{1%
}{3}
\end{equation}%
in the direct sum of the spin space $S=0\oplus S=1\oplus S=2$. The effective
spin-exchange interaction
\begin{eqnarray}
J_{i} &=&2N!\left( \frac{\hbar ^{2}}{M}\right) ^{2}\int dx_{1}\cdots
dx_{N}\left\vert \frac{\partial \phi _{F}}{\partial x_{i}}\right\vert ^{2}
\notag \\
&&\times \delta (x_{i}-x_{i+1})\theta \left( x_{1}\cdots x_{N}\right)
\label{Ji}
\end{eqnarray}%
depends on the overlap between the wave functions of neighboring atoms. The
structure of the Hamiltonian takes the form of a HBB spin-$1$ chain \cite%
{Yip2003,Affleck}
\begin{equation}
H=\sum_{i}\left( \mathbf{s}_{i}\cdot \mathbf{s}_{i+1}+\beta (\mathbf{s}%
_{i}\cdot \mathbf{s}_{i+1})^{2}\right) .  \label{bilinear}
\end{equation}%
The only difference is that here the coupling constants of neighbor spins
are different, due to the background trap potential and the resultant
inhomogeneous particle density. The effective spin Hamiltonian $H_{eff}$,
constructed from variational approach and perturbation theory \cite%
{Deuretzbacher,Yang,Yang1,Lindgren}, conserves the square of the total spin
operator $\mathbf{S}=\sum_{i=1}^{N}\mathbf{s}_{i}$, its $z$ component $%
S_{z}=\sum_{i=1}^{N}s_{i}^{z}$, and the parity operator $\Pi
=P_{1,N}P_{2,N-1}\cdots $, such that the eigenstates of $H_{eff}$ can be
classified in terms of the three quantum numbers: the total spin $S$, the
total magnetization $S_{z}$ and the parity $\Pi $.

\begin{figure}[tbp]
\includegraphics[width=0.50\textwidth]{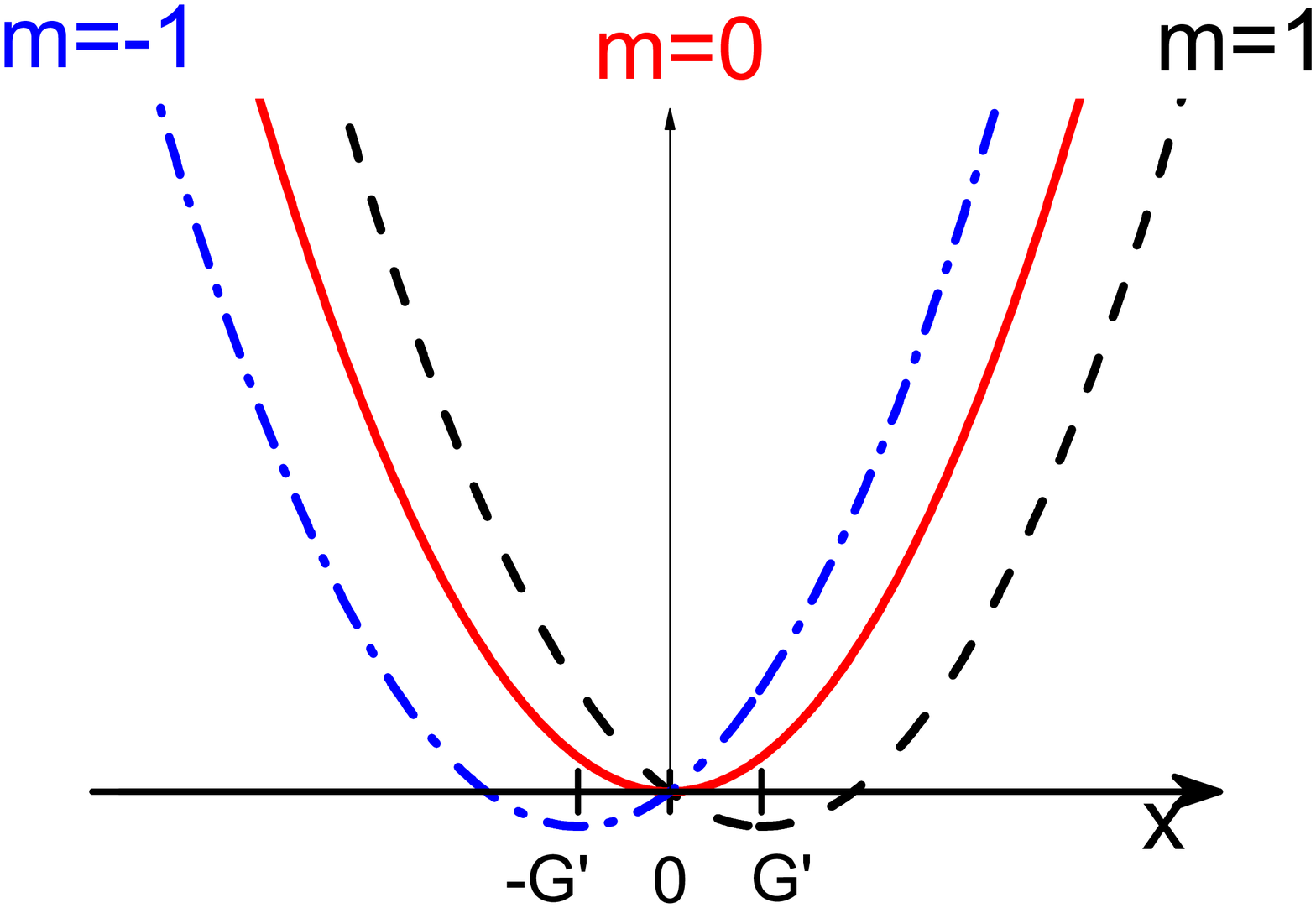} %
\includegraphics[width=0.50\textwidth]{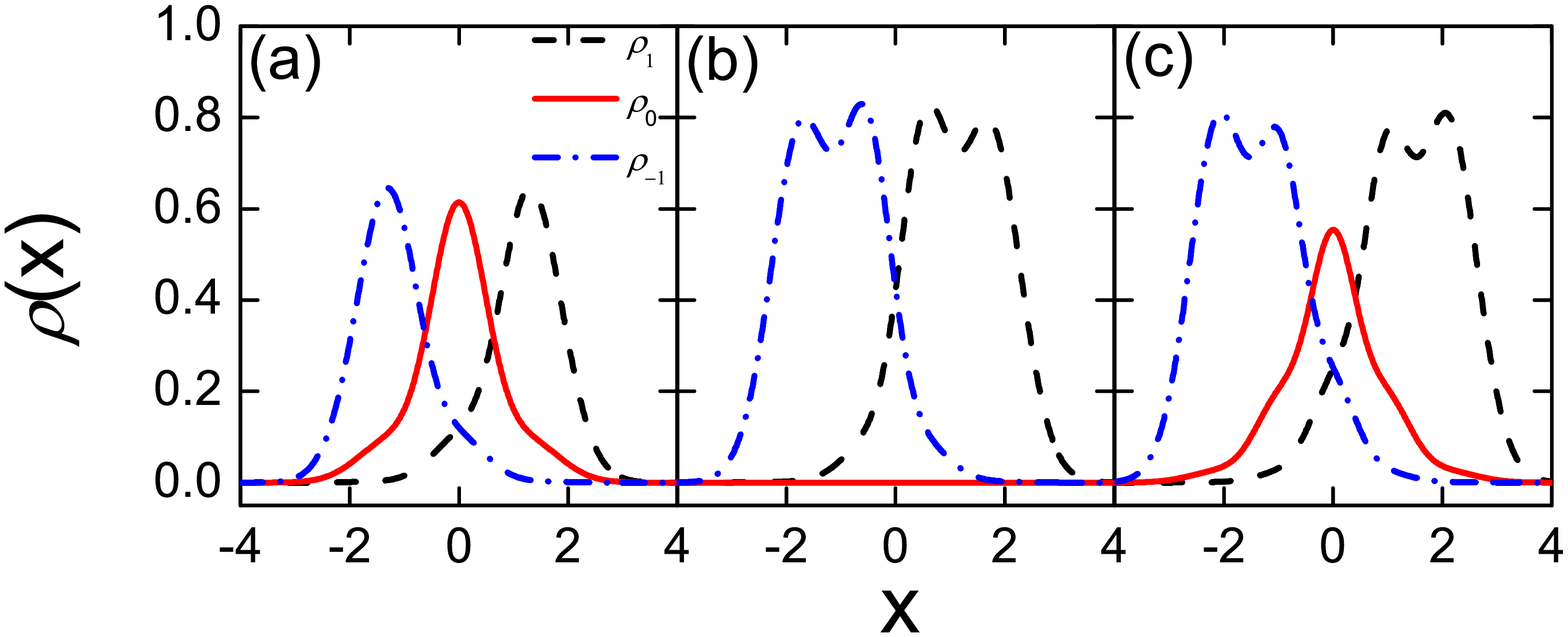}
\caption{(Color online). Potential wells (top) and density distributions
(bottom) for $m=-1$ (blue dot-dashed), $0$ (red solid) and $1$ (black
dashed) in subspace $S_{z}=0$ for a small value of displacement $G^{\prime }$
with $G=2\hbar ^{2}\protect\omega ^{2}/g_{2}$ and $g_{0}/g_{2}=1$ obtained
from the effective spin model. (a) $N=3$; (b) $N=4$; and (c) $N=5$. The
units of the coordinate $x$ and the interaction strength $g_{2}$ are $a_{ho}=%
\protect\sqrt{\hbar /M\protect\omega }$ and $\protect\sqrt{\hbar ^{3}\protect%
\omega /M}$, respectively.}
\label{fig2}
\end{figure}

It is intuitive to examine first the eigenvalues and eigenstates of the $%
H_{eff}$ in the regime of large but finite repulsion $g_{0},g_{2}\gg 0$. In
the simplest case of two particles $N=2$, we can easily see that the
eigenvalues in the channel $S=2,1,0$ are respectively $-J_{1}/g_{2}$, $0$
and $-J_{1}/g_{0}$. While in the anti-ferromagnetic spin chain people pay
more attention to the degenerate point of singlet and triplet $\beta =-1/3$,
which corresponds to TG limit $g_{0}\rightarrow +\infty $ in our case, we
focus on the $SU(3)$ integrable point $g_{0}=g_{2}$ where the quintuplet and
the singlet have the same energy for the ferromagnetic spin chain. Note that
the sign of $J$, hence the order of the energy levels, is inverted for these
two cases, which gives different level crossing point for the ground state.
In Figure {\ref{fig1}}, we show the energy level dependence on the ratio $%
g_{0}/g_{2}$, in which each level is $2S+1$-fold degenerate in the total
spin $S$ channel. We find there exist generally plenty of level crossing in
the degenerate point which can be classified into different bunches. To
specify them, one needs to denote the eigenstates as $\left\vert
E_{S}^{n},S_{z},\Pi \right\rangle $, where $n$ labels the bunch of
degenerate states occurring at $g_{0}=g_{2}$. The two-particle eigenstates
with zero magnetization $S_{z}=0$ can be constructed as
\begin{eqnarray*}
\left\vert E_{2}^{1},0,1\right\rangle &=&\left( \left\vert 1,-1\right\rangle
+2\left\vert 0,0\right\rangle +\left\vert -1,1\right\rangle \right) /\sqrt{6}%
, \\
\left\vert E_{0}^{1},0,1\right\rangle &=&\left( \left\vert 1,-1\right\rangle
-\left\vert 0,0\right\rangle +\left\vert -1,1\right\rangle \right) /\sqrt{3},
\\
\left\vert E_{1}^{2},0,-1\right\rangle &=&\left( \left\vert
1,-1\right\rangle -\left\vert -1,1\right\rangle \right) /\sqrt{2},
\end{eqnarray*}%
on which other states with $S_{z}=\pm 2,\pm 1$ can be obtained by applying
spin raising or lowering operators $S^{\pm }$ repeatedly. We note that two
of them belong to the first bunch $n=1$, while the second bunch $n=2$
consists of a single level. For the three-particle case with $S_{z}=0$,
seven levels group into four bunches with the number of levels 2,2,2,1 in
each bunch respectively; in the ground state bunch, the total spin $S=3$
state with energy $-2J_{1}/g_{2}$ competes with the $S=1$ state with energy $%
-J_{1}/2g_{2}\left( 1/2+2\alpha +\sqrt{\left( 2\alpha -1\right) ^{2}+5/4}%
\right) $, giving rise to the level crossing point at $\alpha =1$ where $%
\alpha =\left( 2g_{2}/g_{0}+1\right) /3$.

The spectrum of $H_{eff}$ for more particles $N=4,5$ are shown in Figure \ref%
{fig1}(c)-(d) by numerically diagonalizing the Hamiltonian (\ref{he}) in the
spin Fock state vector $\left\vert m_{1}m_{2}\cdots m_{N}\right\rangle $. It
is clearly seen that the spectrum is asymmetric about the integrable point $%
g_{0}=g_{2}$, at which the ground state is $\left( N+1\right) \left(
N+2\right) /2$ fold degenerate. The levels belonging to the same bunch have
the same parity: it is always even ($\Pi =1$) for the ground state bunch,
which contains nevertheless $(N+1)/2$ levels for odd $N$ and $(N+2)/2$
levels for even $N$, whereas it is always odd ($\Pi =-1$) for the first
excited state bunch with $(N-1)$ levels. The total spin in the ground state
bunch are $S=N,N-2,N-4,\cdots $ with step $\Delta S=2$ and for the first
excited state $S=N,N-1,\cdots 1$ with step $\Delta S=1$. The two levels with
highest total spin $S=N$ and lowest $S=0(1)$ for even (odd) $N$ are the only
two candidates for the ground state configuration as a consequence of the
ferromagnetic or anti-ferromagnetic coupling of spins. For FM coupling, $%
g_{0}/g_{2}>1$, away from the level crossing point, the degenerate energy
levels of the bound states are eliminated for different total spin $S$. For
every bunch of degenerate energy levels, the energy decreases with total
spin $S$, i.e., $E_{S_{1}}^{n}<E_{S_{2}}^{n}$ when $S_{1}>S_{2}$. Therefore,
the state with $S=N$ is the ground state with completely symmetric spin wave
functions. For AFM coupling $g_{0}/g_{2}<1$, on the other hand, the energy
increases with total spin $S$, i.e., $E_{S_{1}}^{n}<E_{S_{2}}^{n}$ when $%
S_{1}<S_{2}$. Therefore the states with $S=1$ (if $N$ is odd) or $S=0$ (if $%
N $ is even) have the lowest energy. The emergence of level crossing in the
lowest bunch of the energy levels at $g_{0}=g_{2}$ clearly indicates a first-order
transition between AFM and FM phases. For $N$ particles, there exist altogether $[N/2]$ 
independent inhomogeneous spin coupling parameters $J_i$, with the only result being
a slight modification of the energy levels compared with the spectrum without the trapping 
potential, which nevertheless does not change the ordering and crossing of the levels. 

\begin{figure}[tbp]
\includegraphics[width=0.45\textwidth]{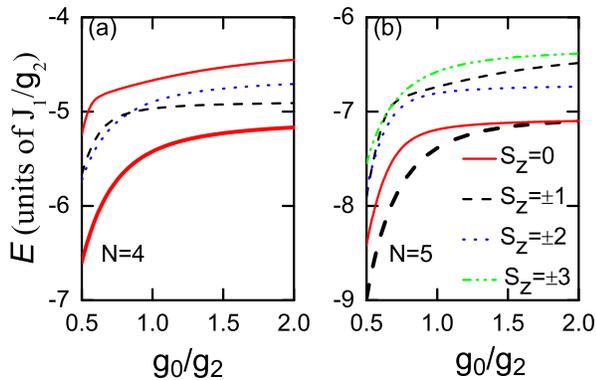}
\caption{The lowest few energy levels of $N=4$ (a) and $N=5$ (b) classified
by $S_{z}$ as a function of $g_{0}/g_{2}$ for a gradient $G=2\hbar ^{2}%
\protect\omega ^{2}/g_{2}$. Clearly the lowest level for $N=4$($N=5$) is $%
S_{z}=0$($S_{z}=\pm 1$).}
\label{fig3}
\end{figure}

The spectrum of the system is highly degenerate for the total spin $S$. We
now consider the weak spin-dependent magnetic gradient introduced in the
single-particle Hamiltonian (\ref{h}). As schematically shown in Fig. \ref%
{fig2}, atoms of different spin components are trapped in different
potential wells, with the trap center moved to the left or right by an
amount $G^{\prime }=G\hbar /M\omega ^{2}$ depending on the value of spin $m$%
. The corresponding effective spin Hamiltonian $\left( \ref{he}\right) $ in
the limit of strong interaction will be modified into
\begin{equation}
H_{eff}^{\prime }=H_{eff}-G\sum_{i}D_{i}s_{i}^{z},  \label{hg}
\end{equation}%
where $D_{i}=N!\int x_{i}\left\vert \phi _{F}\right\vert ^{2}\theta \left(
x_{1}\cdots x_{N}\right) \prod\nolimits_{j=1}^{N}dx_{j}$ represents the
average position of the $i$th atom. The spin-dependent magnetic gradient
destroys the total spin conservation and parity conservation, implying that $%
H_{eff}^{\prime }$ no longer commutes wtih $S^{2}$ and $\Pi $. However we
can find that the Hamiltonian $H_{eff}^{\prime }$ commutes with an operator $%
T=\Pi \prod_{j=1}^{N}a_{j}$ \cite{ChenLi}, where
\begin{equation}
a_{j}=\left(
\begin{array}{ccc}
0 & 0 & 1 \\
0 & 1 & 0 \\
1 & 0 & 0%
\end{array}%
\right)
\end{equation}%
serves to flip the spin of $j$-th atom. It is straightforward to show that $%
\left\{ T,S_{z}\right\} =0$ and $\left[ T,S_{z}\right] =2TS_{z}$. As a
result, $T$ applying to an energy eigenstate $\left\vert
E_{i},S_{z}\right\rangle $ changes the state to a degenerate eigenstate $%
\left\vert E_{i},-S_{z}\right\rangle $, i.e.,
\begin{equation}
T\left\vert E_{i},S_{z}\right\rangle =\left\vert E_{i},-S_{z}\right\rangle .
\end{equation}%
One can also infer from Fig. \ref{fig3} that, adding the magnetic gradient
lifts partially the degeneracy of spectrum. The states with total
magnetizations $S_{z}$ and $-S_{z}$ remain degenerate. The ground state of
the system mixes all total spin states to achieve a maximum reduction in
energy and thus occurs the spin component separation (bottom of Fig. \ref%
{fig2}). The first-order transition at $g_{0}=g_{2}$ disappears and the
level with total magnetization $S_{z}=0$ proves to be the ground state (Fig. %
\ref{fig3}(a)) for even particle numbers. On the other hand, for odd
particle numbers, the state with total magnetization $S_{z}=\pm 1$ has the
lowest energy (Fig. \ref{fig3}(b)).

\begin{figure}[tbp]
\includegraphics[width=0.45\textwidth]{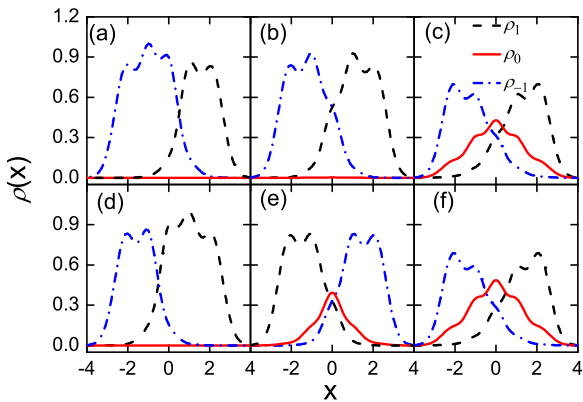}
\caption{Effect of transverse magnetic field and non-integrable interaction
on the density distributions of the ground state for spin component $1$
(black dashed), $0 $ (red solid) and $-1$ (blue dot-dashed) for $N=5$ and $%
G=2\hbar ^{2}\protect\omega ^{2}/g_{2}$. $g_{0}/g_{2}=1$ for (a)-(d) and $%
g_{0}/g_{2}=2$ for (e) and (f). (a) and (d) are two degenerate states of $%
S_{z}=\pm 1$ with $\Omega =0$. $\Omega =0.001\hbar \protect\omega $ for (b)
and (e), $\Omega =0.05\hbar \protect\omega $ for (c) and (f). }
\label{fig4}
\end{figure}

\begin{figure*}[tbp]
\includegraphics[width=0.2\textwidth]{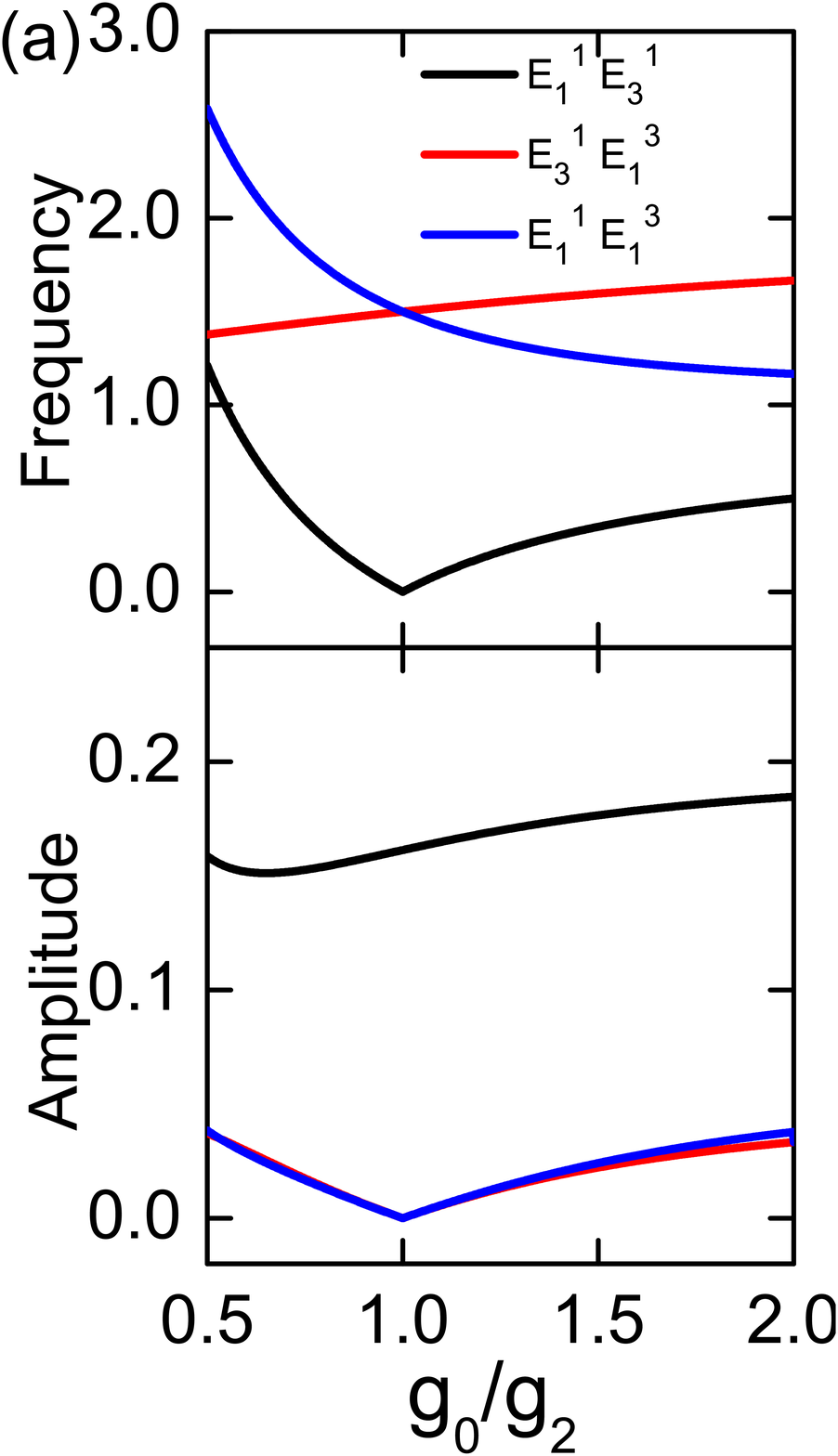} \includegraphics[width=0.4%
\textwidth]{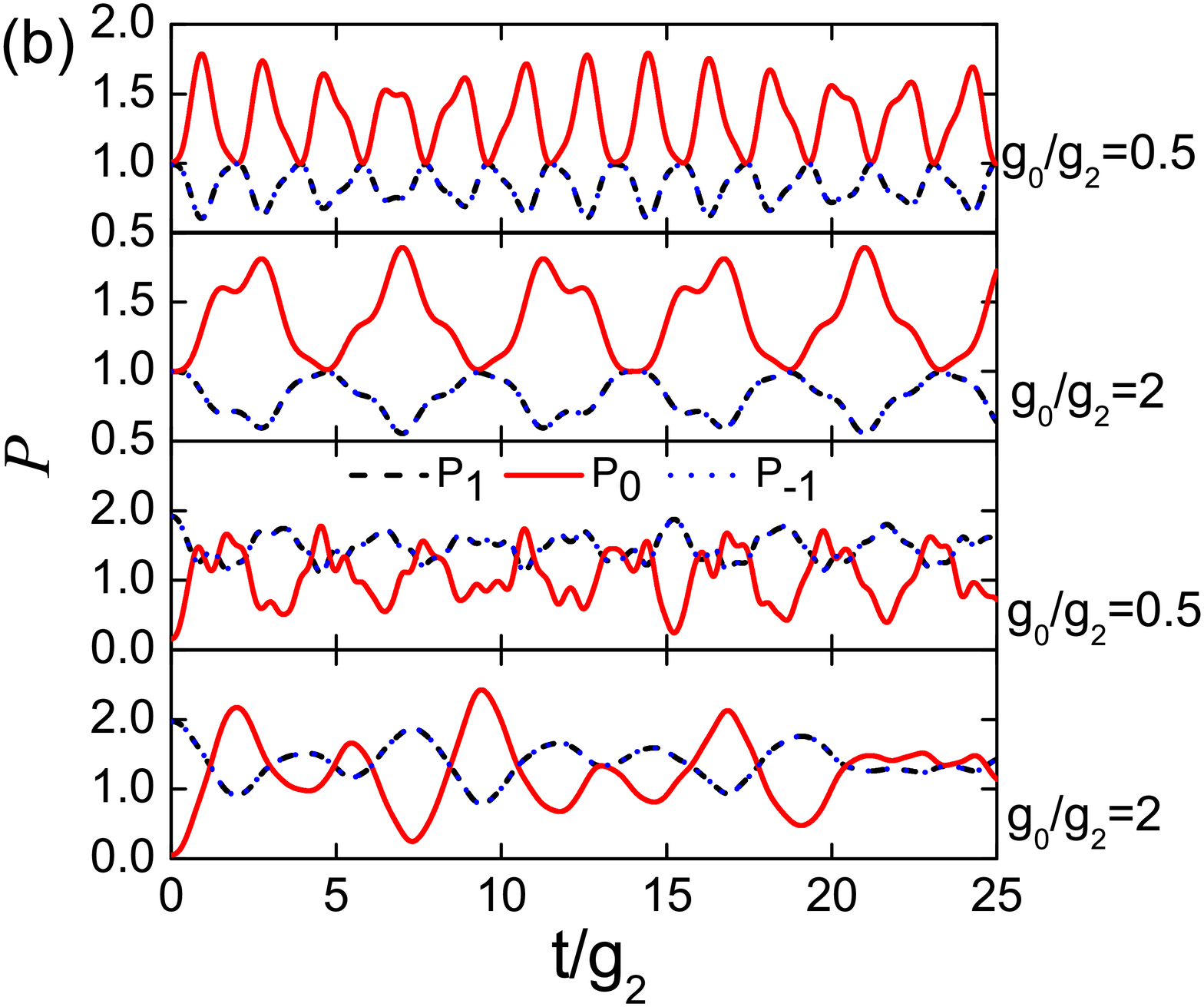}
\caption{Frequencies and amplitudes of the population of spin $1$ or $-1$
component versus $g_{0}/g_{2}$ for $N=3$ (a) and the populations
oscillations $P_{m}$ of spin component $m=0,\pm1$ as a function of $t/g_{2}$
for $g_{0}/g_{2}=2$ and $g_{0}/g_{2}=0.5$. The frequency is in units of $%
J_1/2\protect\pi\hbar g_2$ and the unit of $t/g_{2}$ is $\protect\sqrt{%
m/\hbar ^{3}\protect\omega ^{3}}$. The top two panels are for $N=3$ and the
bottom two are for $N=4$. The black dashed curves show the population of $%
m=1 $, the red solid curves show the population of $m=0$ and the blue dotted
curves show the population of $m=-1$.}
\label{fig5}
\end{figure*}

The density distribution of $m$-th spin component is defined as
\begin{equation}
\rho_m (x) = \sum_i \rho_m^{(i)} \rho^{(i)}(x)
\end{equation}
with the probability that the magnetization of the $i$th spin equals $m$,
\begin{equation}
\rho _{m}^{\left( i\right) } =\sum_{m_{1},\cdots,m_{N}}\left\vert
\left\langle m_{1},\cdots ,m_{N}|\chi \right\rangle \right\vert ^{2}\delta
_{m,m_{i}}  \label{rhomi}
\end{equation}
and the probability to find the $i$th atom with any spin at position $x$,
\begin{equation}
\rho^{(i)}(x)=N! \int dx_1\cdots dx_N \delta(x-x_i) \theta (x_1,\cdots, x_N)
\left\vert\phi_F\right\vert^2.
\end{equation}
We show the density distribution for a gradient $G=2\hbar^2 \omega^2 /g_2$
in the bottom of Fig. {\ref{fig2}}. To be more precise, we focus on the $%
SU(3)$ integrable point $g_0=g_2$ which guarantees the conservation of atoms
in each spin component due to the spin-independent interaction \cite{Jiang}.
Surprisingly, we find that the spin-$0$ component always disappears for even
particle numbers in the subspace of the total magnetization $S_{z}=0$ (see
Fig. \ref{fig2}(b)), while for odd atom numbers the density of spin-0
component remains unity (Fig. \ref{fig2}(a) and \ref{fig2}(c)). This can be
understood by noting that the already fermionized atoms would fill the
evenly spaced levels from bottom one by one, and it is more energetically
favorable to put the additional atoms in the left and right traps which are
lowered by the gradient by an amount $(G\hbar)^2/2M\omega^2$.

For an applied transverse magnetic field the effective Hamiltonian $%
H_{eff}^{\prime }$, with an additional term $\Omega S_x$ included, no longer
commutes with $S_{z}$ and the degeneracy of the system is completely
eliminated. However, $H_{eff}^{\prime }$ still commutes with the operator $T$%
. For even atom numbers, the density distributions in the ground state (see
for example, Fig. \ref{fig2}(b)) are hardly modified after the introduction
of a very small $\Omega $, which can be regarded as a perturbation to the $%
S_{z}=0$ ground state. For odd atom numbers, the ground state can be
constructed as the superposition of two degenerate ground states with $%
S_{z}=\pm 1$, whose densities are shown in Fig. \ref{fig4}(a) and \ref{fig4}%
(d), respectively, i.e.
\begin{equation}
\left\vert G_{odd} \right\rangle= \frac {1}{\sqrt{2}}\left(\left \vert
E_0,S_z=+1 \right \rangle+\left \vert E_0,S_z=-1 \right \rangle \right),
\end{equation}
with $T\left\vert G_{odd} \right\rangle=\left\vert G_{odd} \right\rangle$.
The transverse field plays the role of coupling the two degenerate states
such that the ground state is lowered by an amount $\hbar\Omega$. The
density profiles of spin $+1$ and spin $-1$ are now symmetric to each other,
due to the conservation of $T$ implying an combined operation of space
inversion and spin flipping, as can be seen in Fig. \ref{fig4}(b). The
population in the component spin $0$ increases noticeably for a strong
enough $\Omega $ (Fig. \ref{fig4}(c)). The off-diagonal feature of the $S_x$
matrix would inevitably mix excited states such as that with $S_z =0$, the
spin $0$ density of which is significant. An alternative way to introduce
the spin $0$ component is to bring the system away from the integrable
point. We show this in Fig. \ref{fig4}(e) and \ref{fig4}(f) for the case of $%
g_0/g_2=2$, which shows an obvious enhancement of spin $0$ density.

\section{Dynamics}

\subsection{Spin-changing Dynamics}

Realizing spin-chain Hamiltonian with trapped cold atoms offers important
applications in the study of microscopic magnetic phenomena. Here we
investigate the spin-changing dynamics of this system, which is different
from the population dynamics of the weakly interacting system governed by
the Gorss-Pitaevskii equation \cite{Chang,Kronjager}. To do this, strongly
interacting atoms are initially prepared in the ground state $\left\vert
\chi \left( 0\right) \right\rangle $ with a weak spin-dependent magnetic
gradient $G=2\hbar ^{2}\omega ^{2}/g_{2}$. The total magnetization $S_{z}$
is still the conserved quantity, so the system will evolve within one of the
$S_{z}$ subspace. Then the gradient $G$ is abruptly switched off and the
evolution of the system is governed by the effective spin chain Hamiltonian $%
H_{eff}$ in (\ref{he}). The initial state is realized by obtaining the
ground state of Hamiltonian $H_{eff}^{\prime }$ with non-vanishing $G$.
Starting from the initial state $\left\vert \chi \left( 0\right)
\right\rangle $, the time evolution of the wave function is governed by%
\begin{equation*}
\left\vert \chi \left( t\right) \right\rangle =e^{-\frac{i}{\hbar }%
H_{eff}t}\left\vert \chi \left( 0\right) \right\rangle =\sum_{i}c_{i}e^{-%
\frac{i}{\hbar }E_{i}t}\left\vert \phi _{i}\right\rangle ,
\end{equation*}%
where $c_{i}=\left\langle \phi _{i}\right. \left\vert \chi \left( 0\right)
\right\rangle $ is the overlap of the initial state and the $i$-th
eigenstate of the system $\phi _{i}$ with eigenenergy $E_{i}$. We introduce
the spin population $P_{m}\left( t\right) =\sum_{i}\rho _{m}^{\left(
i\right) }\left( t\right) $ with $\rho _{m}^{\left( i\right) }\left(
t\right) $ defined in (\ref{rhomi}) with the replacement $\chi
(0)\rightarrow \chi (t)$, which measures the population of $m$-th component
in the system. For spin-$1$ system, two atoms in the states $-1$ and $+1$
have a chance to coherently and reversibly scatter into final states
containing two atoms in the state $0$, which leads to the population
transferring from $P_{1}\left( t\right) +P_{-1}\left( t\right) $ to $%
2P_{0}\left( t\right) $, or vice versa, subject to the conservation of the
total population $\sum_{m}P_{m}\left( t\right) =N$. The system satisfies $%
P_{1}\left( t\right) -P_{-1}\left( t\right) =S_{z}$ at any time. In this
section we only consider the dynamics in the subspace of total magnetization
$S_{z}=0$ in which case one must have $P_{+1}=P_{-1}$. We illustrate the
spin population dynamics in Figure \ref{fig5} for both $P_{\pm 1}$ and $%
P_{0} $. In the case of $N=3$, the initial spin populations for both
interaction parameters $g_{0}/g_{2}=0.5$ and $2$ are very close to the case
of equally distributed among the three components in the integrable point $%
g_{0}=g_{2}$. Starting from such an initial population, Rabi-like
oscillations of spin populations between the components $0$ and $\pm 1$ are
observed and depicted in Fig. \ref{fig5}(b), which is in sharp contrast to
the respectively conservation of atoms in each spin component for $%
g_{0}=g_{2}$. In the entire range of interaction of interest, we managed to
extract the amplitude and the frequency of the oscillation (Fig. \ref{fig5}%
(a)), and it turns out that the oscillation amplitudes of populations are
determined by the weight coefficients of the basis vectors and the
oscillation frequencies of populations are determined by the energy
differences, among which three energy levels $E_{1}^{1},E_{3}^{1},E_{3}^{3}$
play dominate roles in the dynamics of the spin-changing collisions. At the
integrable point we find either the frequency or the amplitude of the
partial wave would vanish, which ensures the populations in each components
remain constants, $P_{0}(t)=P_{\pm 1}(t)=1$ for $N=3$ and $P_{0}(t)=0$, $%
P_{\pm }(t)=2$ for $N=4 $. The intrinsic origin of this exotic phenomenon
lies in that this point is highly degenerate. Away from this point, the
oscillation frequency of the primary amplitude increases significantly on
both sides of $g_{0}=g_{2}$, however, a lower frequency will slow down the
oscillation for $g_{0}/g_{2}=2$. More energy levels are involved in the
dynamics of $N=4$ atoms, the initial spin population of which is close to
the case of equally distributed on the $\pm 1$ components at $g_{0}=g_{2}$.
The characteristic dynamics here may be used to detect the quantum phases of
the spin-1 chain model, and moreover, may reveal the interesting spin
population transfer across the phase boundary by the oscillation frequency.

\begin{figure}[tbp]
\includegraphics[width=0.45\textwidth]{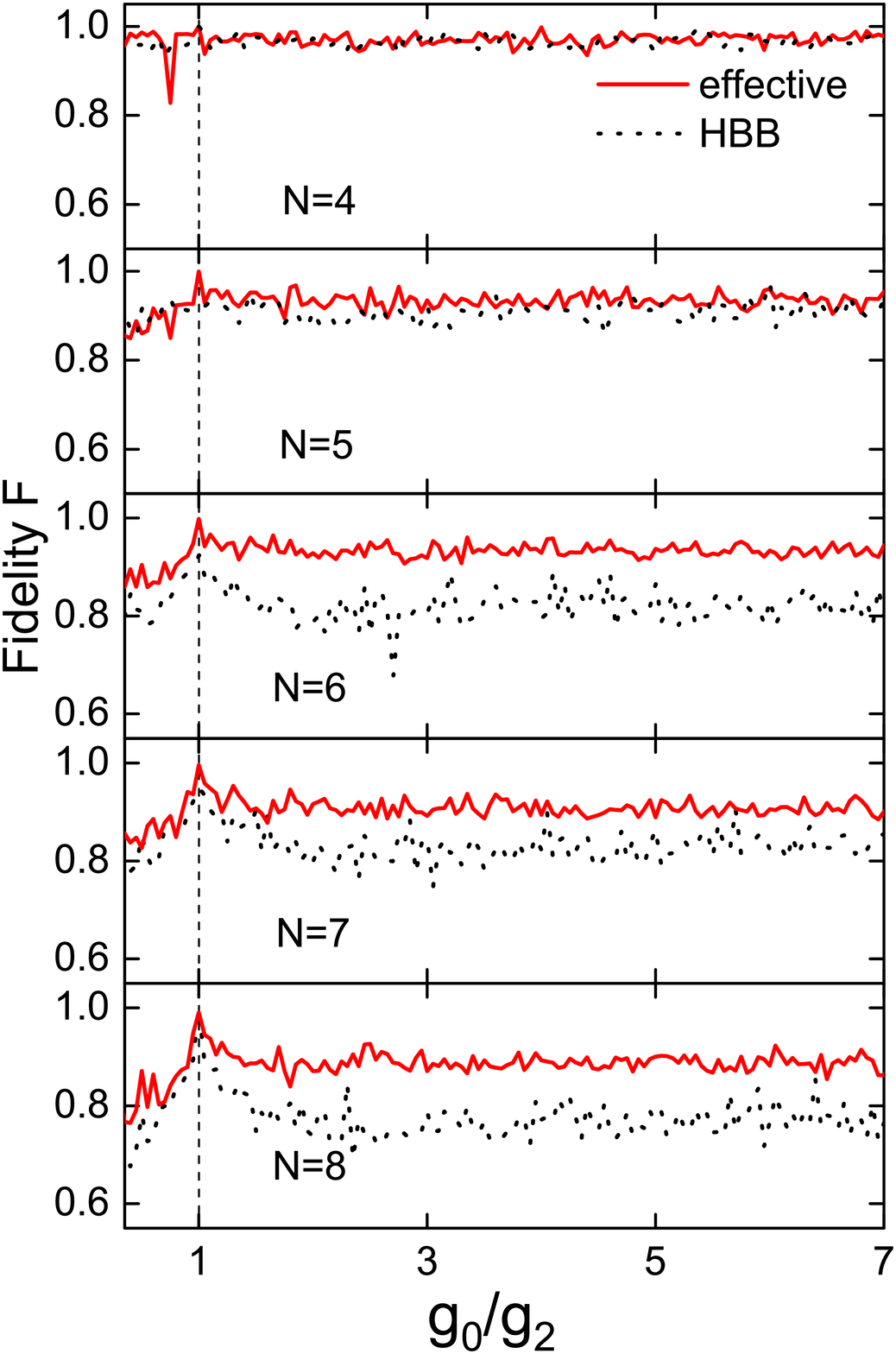}
\caption{(color online) Maximum value of the fidelity $F$ of state transfer
in a spin-$1$ system with particle numbers $N=4$ to $8$, for the HBB spin
chain (black dotted) and the effective spin chain (red solid). Vertical
dotted line indicates the phase transition point $g_{0}/g_{2}=1$.}
\label{fig6}
\end{figure}

\subsection{Quantum state transfer}

Spin chains have important applications in quantum simulation and
computation. The spin chains have been proposed intensively as quantum
channels to study state transfer in small quantum networks \cite%
{Bose2003,Bose2007,Burgarth}. Perfect quantum-state transfer is very
important to accomplish prospective quantum information processing through a
chain of nearest-neighbor coupled spins. The interaction energy of each
qutrit-qutrit pair in the translation-invariant HBB spin-$1$ chain is the
same, which can be described by the Hamiltonian (\ref{bilinear}) with $\beta
=(2g_{2}/g_{0}+1)/3$, but in our effective spin chain this interaction
energy (\ref{Ji}) is site dependent. Here we study the superiority of the
inhomogeneous effective spin-1 chain as a quantum channel.

Transferring a known or unknown quantum state with spin-1 from one place to
another has been studied in Ref. \cite{Ghosh,Wiesniak}. In the original
proposal, the quantum-state transfer protocol involves initializing the spin
chain of $N$ sites with the first spin in an arbitrary state $\left\vert
\psi \right\rangle =\xi _{-1}\left\vert -1\right\rangle +\xi _{0}\left\vert
0\right\rangle +\xi _{1}\left\vert 1\right\rangle $ ($\sum_{m}\left\vert \xi
_{m}\right\vert ^{2}=1$) and decoupled from the rest of the chain. At $t=0$,
the first and second spins abruptly couple and let the system freely evolve
in the spin chain. At time $t$, the quality of the transfer of $\left\vert
\psi \right\rangle $ to the last spin of the chain is evaluated by the
fidelity of attaining $\left\vert \psi \right\rangle $ at site $N$. Ideal
transfer would imply that at time $t^{\ast }$ the last spin of the chain is
in state $\left\vert \psi \right\rangle $. We consider a simple case, at
time $t=0$\ its state was $\left\vert \Psi \left( 0\right) \right\rangle
=\left\vert -1,1\cdots 1\right\rangle $. Our aim is to maximize the
probability of retrieving state $\left\vert 1\cdots 1,-1\right\rangle $\ at
time $t^{\ast }$. We define the fidelity of state transfer as
\begin{equation*}
F\left( t\right) \equiv \left\vert \left\langle \Psi \left( t\right)
|1\cdots 1,-1\right\rangle \right\vert ^{2},
\end{equation*}%
which relies only on the expansion coefficients of the eigenstates $%
\left\vert E_{S}^{n}\right\rangle $ expanded in the basis vectors $%
\left\vert 1\cdots 1,-1\right\rangle $ and $\left\vert -1,1\cdots
1\right\rangle $. Let $F\equiv F\left( t^{\ast }\right) $ be the maximum
value that achieves in the intermediate time. We plot the maximum of
fidelity $F$ of the state transfer in Fig. \ref{fig6} as a function of $%
g_{0}/g_{2}$. For a system of $N=4$ with the trap frequency $\omega =40kHz$
and the interaction strength $g_{2}=20$ in units of $\sqrt{\hbar ^{3}\omega
/M}$, the occurrence time of the maximal fidelity $t^{\ast }$ ranges from $%
10^{-2}$s to $1.5$s, which appears, however, randomly for varying $%
g_{0}/g_{2}$. We compare the fidelity $F$ of the HBB and effective spin
chains of length $N=4$ to $8$ in Fig. \ref{fig6}, which reflects that the
effective spin chain transfers the state more faithfully than the HBB spin
chain, especially for longer spin chain. With the increase of particle
number $N$, the overall trend is that $F$ decreases. At the integrable point
$g_{0}/g_{2}=1$, $F$ reaches a maximum value in the both spin chain models,
while the effective spin chain model always provides more efficient way for
quantum state transfer in the entire interaction regime.

\section{Conclusions}

We have shown that a three-component system of strongly interacting bosonic
atoms in a 1D harmonic trap can be represented effectively as a spin chain
described by the bilinear-biquadratic spin-1 model Hamiltonian. For few
atoms in the trap we have determined the energy spectrum of the ground
states and obtained the rules of the ordering and crossing of energy levels
near the first-order quantum phase transition, i.e. the $SU(3)$ integrable
point, $g_0=g_2$. The energy levels of the eigenstates are collected into
different bunches which can be labelled by the total spin and the parity.
Away from the degenerate point, the ground state is either with highest
total spin $S=N$ for FM coupling between atoms, or with lowest spin $S=1$
(for odd $N$) or $S=0$ (for even $N$) for AFM coupling $g_0/g_2 <1$. We
further introduce a magnetic gradient to remove the degeneracy on $S$,
motivated by the experimental studies of coherent multi-flavour spin
dynamics in a fermionic quantum gas \cite{Krauser}, and subsequently study
the quench dynamics of the the ground states of spin component separation
when the initial magnetic gradient is removed quickly. Our results reveal
the spin-change dynamics of the system governed by the ratio of interactions
between the two channels. Through the study of the dynamics of the quantum
state transfer, we show that the inhomogeneous qutrit-qutrit interaction of
the engineered effective spin chain is more efficient in state transferring.

\begin{acknowledgments}
Y. Zhang would like thank Han Pu for helpful discussion. This work is
supported by NSF of China under Grant Nos. 11474189, 11674201, 11234008,
Program for Changjiang Scholars and Innovative Research Team in University
(PCSIRT)(No. IRT13076). S. C. is supported by NSF of China under Grant Nos.
11425419, 11374354 and 11174360.
\end{acknowledgments}

\end{document}